\documentclass[prl,twocolumn,amsmath,amssymb,groupedaddress]{revtex4}

\usepackage{dcolumn}
\usepackage{graphicx,subfigure}
\usepackage{bm}
\usepackage{verbatim}
\usepackage{amsmath}
\usepackage{amssymb}
\usepackage[T1]{fontenc}
\usepackage{ae,aecompl}
\usepackage{appendix}
\usepackage{float}
\usepackage{color}

\newcommand{\be}{\begin{equation}}
\newcommand{\ee}{\end{equation}}
\newcommand{\bea}{\begin{eqnarray}}
\newcommand{\eea}{\end{eqnarray}}
\newcommand{\bse}{\begin{subequations}}
\newcommand{\ese}{\end{subequations}}
\def\rf#1{(\ref{#1})}

\begin{document}
\title{Reentrant supersolidity}
\author{Leo Radzihovsky}
\affiliation{
Department of Physics and Center for Theory of Quantum Matter\\
University of Colorado, Boulder, CO 80309}
\date{December 2015}
\email{radzihov@colorado.edu}

\begin{abstract}
  A ``supersolid'' -- a crystal that exhibits an off-diagonal
  long-range order and a superflow -- has been a subject of much
  research since its first proposal\cite{AndreevLifshitz69}, but has
  not been realized as a ground state of short-range interacting
  bosons in a continuum.  In this note I point out a simple and
  generic mechanism for a thermally-driven reentrant supersolidity,
  and discuss challenges of  experimental realization of this
  idea. In the limit of bosons in a periodic potential, this mechanism
  reduces to  a {\em reentrant} low-temperature
  normal-superfluid transition, that should be accessible to
  simulations and in current experiments on bosonic atoms in an optical
  periodic potential.
\end{abstract}

\maketitle

\noindent{\em Introduction and background:}
A ``supersolid'' -- a crystal that can superflow is a putative state
of matter that spontaneously ``breaks'' global translational and
$U(1)$ phase rotational symmetries, associated with momentum and
particle number conservation,
respectively.\cite{AndreevLifshitz69,Chester70} This novel
``incommensurate crystal'' \footnote{In an incommensurate crystal the
  number of lattice sites is decoupled from the number of atoms.}
ground state is driven by quantum fluctuations that can lower the
kinetic energy due delocalization of a fraction of atoms from their
lattice sites below the interaction energy cost of vacancies and
interstitials, i.e., a vanishing of the Mott gap while retaining
crystalline order. Such putative state has been somewhat of a holy
grail in experiments\cite{KimChanNature04, KimChanScience04,
  RittnerReppy06, KetterleNature17, Fraass89} and
simulations\cite{MeiselReviewSS92, Ceperley04, ProkofevSvistunov05},
as a ground state of short-range interacting bosons in a
continuum.\footnote{A generalized supersolid with its definition
  expanded to include bosons with dipolar and/or spin-orbit
  interactions, vortex lattices of rotated condensates, and lattice
  bosons in a periodic potentials with a fractional filling have
  indeed been realized, and many others, like the
  Fulde-Ferrel-Larkin-Ovchinikov and Pair-Density Wave superconductors
  as promising candidates.} However, despite these promising but
controversial experiments\cite{KimDyChanPRB14}, that rekindled this
seasoned subject, to my knowledge for a continuous translational
symmetry no conclusive observation exist to date.

Despite this currently discouraging experimental status, in this note
I point out a simple generic mechanism -- with its own challenges,
that I will discuss -- for a temperature-tuned reentrant supersolid
state, that should be of interest even if a zero-temperature
supersolid is not realized.  One way to characterize the supersolid
state is as a superfluid of vacancies and/or interstitials that are
driven by quantum fluctuations to proliferate even in the ground state
($T=0$) of a quantum crystal of bosonic atoms. Such putative state
will then generically be stable at low temperatures, but will
transition into a conventional crystal when at high temperature
vacancies and interstitials undergo a phase transition into a normal
liquid or a thermal Boltzmann gas, as any gas inevitably does.

\noindent{\em BEC of vacancies/interstitials - a reentrant
  supersolid:}
As in a conventional Bose-condensate (BEC), a phase transition to a
BEC (SF) state of vacancies and interstitials is controlled by the
dimensionless phase-space density $n\lambda_T^3$ exceeding a constant
of order $1$. This corresponds to the condition that the
three-dimensional (3D) thermal de Broglie wavelength
$\lambda_T = \sqrt{2\pi\hbar^2/m T}$ (using units with $k_B= 1$)
exceeds inter-particle spacing $n^{-1/3}$.  Equivalently, it is useful
to state the criterion for a BEC transition in terms of a critical
boson density,
\begin{equation}
  n_c(T) = \left(2m T/\hbar^2\right)^{3/2},
  \label{nc}
\end{equation}
with BEC emerging for $n > n_c(T)$ (up to a constant of order 1).

Somewhat counter-intuitively, here I consider a crystal ground state,
which at zero temperature is a conventional ``commensurate''
(nonsuper-) solid, and focus on its fate upon warming it to a nonzero
temperature.  I then note that there are two competing effects of
increasing temperature of such a state. On one hand, as usual,
increasing temperature decoheres a quantum state, reducing particles'
thermal de Broglie wavelength.  On the other hand, since vacancies and
interstitials in a crystal are point defects with a finite core energy
$E_c$, a finite density of them, $n(T)$ is induced at any nonzero
temperature through thermal activation,
\begin{equation}
  n(T) = n_0 e^{-E_c/T},
  \label{nT}
\end{equation}
where $n_0\sim 1/a^3$ is the total density of bosonic atoms (with $a$
the crystal's lattice constant), or equivalently the density of the
bosonic fluid were the crystal to be melted.  With the vacancies and
interstitials' core energy $E_c$ set by short-scale elastic energies I
expect it to be proportional to the bulk modulus $B$ and roughly given
by
\begin{equation}
E_c \sim a^3 B. 
  \label{Ec}
\end{equation}

\begin{figure}[tbp]
  \includegraphics[height=3.5in,width=3in]{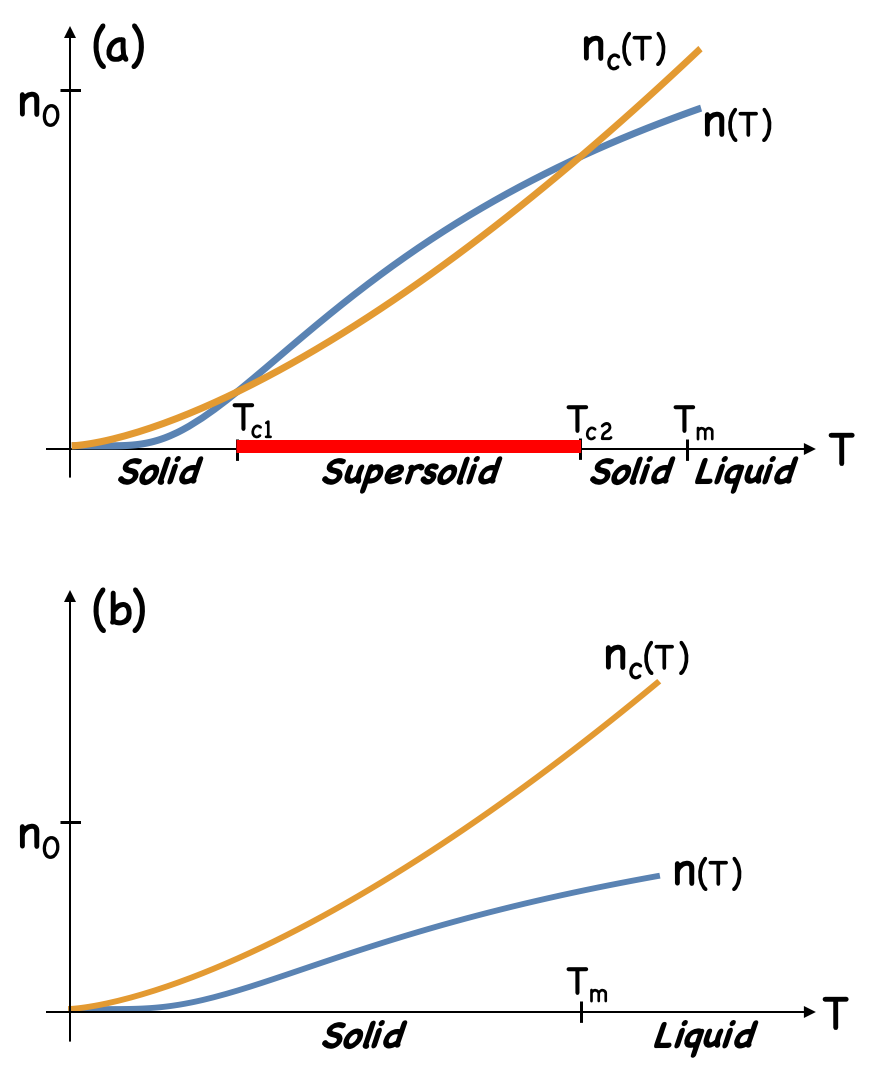}
  \caption{Phase diagrams for a quantum crystal under two scenarios.
    (a) $E_c \ll T_{BEC}(n_0)$, exhibiting reentrant supersolidity,
    where a conventional crystal (Solid) undergoes a phase transition
    at $T_{c1}$ into a Supersolid as thermally-proliferated
    vacancies/interstitials Bose-condense with increasing $T$,
    followed by a reentrant transition at $T_{c2}$ back into a
    conventional crystal (with a thermal gas of
    vacancies/interstitials). The solid then subsequently melts at
    $T_m$. (b) $E_c \gg T_{BEC}(n_0)$, exhibiting a conventional
    crystal (Solid) that melts at $T_m$ directly into a conventional
    fluid (Liquid) phases.}
\label{figRentrance}
\end{figure}

As illustrated in Fig.\ref{figRentrance}, because $n(T)$ exhibits an
essential singularity at zero temperature, plateauing to $n_0$ at high
$T$, while $n_c(T)$ is a power-law as a function of temperature,
generically (without fine-tuning) the two functions are guaranteed to
either not cross for $E_c\gg T_{BEC}(n_0)$ or to cross {\em twice} for
$E_c\ll T_{BEC}(n_0)$.  In the former case (Fig.\ref{figRentrance}(b))
vacancies and interstitials remain too dilute to Bose-condense, with a
conventional crystal appearing at all temperatures below the melting
temperature $T_m$. In the latter case (Fig.\ref{figRentrance}(a))
$n(T)$ and $n_c(T)$ cross at two temperatures $T_{c1}$ and $T_{c2}$
set by $n(T) = n_c(T)$,
\begin{equation}
  n_0 e^{-E_c/T} = \left(2m T/\hbar^2\right)^{3/2},
  \label{crossEq}
\end{equation}
as illustrated in Fig.\ref{figRentrance}(a).

As long as the crystal does not melt this corresponds to a reentrant
supersolidity as a function of temperature,
\begin{eqnarray}
  &&0 < T < T_{c1}, \ \text{conventional crystal S},\\
   &&T_{c1} < T < T_{c2}, \ \text{supersolid SS},\;\;\;\;\;\;\\
 &&T_{c2} < T, \ \text{conventional crystal S}.
\end{eqnarray}
It is straightforward to see that these phase boundaries are set by,
\begin{eqnarray}
  T_{c1} &\sim& E_c,\\
  T_{c2} &\sim& T_{BEC}= \hbar^2 n_0^{2/3}/2m,
\label{Tc1Tc2}
\end{eqnarray}
Thus, for above scenario of $E_c \ll T_{BEC}(n_0)$, a supersolidity
emerges as a result of {\em increasing} temperature, for $T > T_{c1}$.
For $T < T_{c1}$ the density of vacancies and interstitials is too low
to reach the necessary thermal density $n_c(T)$, and so the gas is
thermal and the state is a conventional solid. In the higher
temperature range of $T_{c1} < T < T_{c2}$, $n(T)$ increases beyond
$n_c(T)$ and the gas of vacancies and interstitials Bose-condenses,
and thereby leads to a supersolid state, despite a concomitant
decrease in the thermal deBroglie length. Then, for $T > T_{c2}$, the
thermal deBroglie length gets sufficiently short that $n(T)$ again
drops below $n_c(T)$ leading to a conventional solid with a thermal
gas of vacancies and interstitials, which then can melt at $T_m$ into
a normal fluid.  Another scenario (not illustrated in
Fig. \ref{figRentrance}) is that the melting temperature falls in the
range $T_{c1} < T_m < T_{c2}$, that corresponds to a supersolid
melting into a superfluid, which then transitions into a normal fluid
at $T_{c2}$.

\noindent{\em Challenges to the realization:}
I now turn to deliberations about different physical scenarios and
challenges to their realization.  To this end, I first note that if
the core energy $E_c$ is determined by the Coulomb energy at the
lattice scale, (i.e., on the scale of a Rydberg (Ry)), it is highly
unfavorable for reentrant supersolidity, since then $T_{BEC}(n_0)$ is
estimated to be three orders of magnitude smaller than $E_c$ by the
ratio of an electron to proton mass.

A more favorable condition for the reentrant supersolid is expected if
$E_c$ is controlled by the energetics of the van der Waals interaction
in neutral atoms, orders of magnitude lower than a Ry. This then
allows for the scenario $E_c \ll T_{BEC}(n_0)$, a necessary but not
sufficient condition for the realization of reentrant supersolidity.
Another necessary condition is on the melting temperature,
$T_m \gg E_c$.  If satisfied, indeed a reentrant supersolid will
appear for $T_{c1} < T < T_{c2}$, as illustrated in
Fig.\ref{figRentrance}(a). On the other hand, generically one may
expect that $E_c$, controlled by short-scale solid's cohesion energies
is comparable to the melting temperature, $T_m$. If so then instead of
a rentrant supersolid, for $T_{c1} \sim T_m < T < T_{c2}$ a
low-temperature conventional crystal will simply melt directly into a
superfluid (BEC), which will then become a thermal fluid (Boltzmann
gas) for $T > T_{c2}$.

Currently, in solid Helium-4 there is no experimental evidence for a
thermally-reentrant supersolid in many hundreds of torsional
oscillator experiments typically performed between $30$ mK and $1.5$
K\cite{MeiselReviewSS92,KimChanNature04,KimChanScience04,RittnerReppy06}. This
is consistent with X-ray measurements by Simmons, et
al.\cite{Fraass89} and with Quantum Monte Carlo
simulations\cite{Ceperley04,ProkofevSvistunovTroyer06}, that estimate
$E_c$ on the order of $10-15$K, thus much larger than
$k_B T_{BEC}(n_0)\sim 1$K.

I note that there is no fundamental obstruction for
elastically-controlled core energy $E_c\sim T_{c1}$ to be lower than
the mass- and density-dependent $T_{BEC }(n_0)\sim T_{c2}$. In
contrast, as emphasized above, in a single-component crystal there is
a natural elastic link between $E_c$ and $T_m$.  Thus to realize
$E_c \ll T_{BEC}(n_0) \ll T_m$ requires a system in which
vacancy/interstitial core energy $E_c$ is decoupled from its melting
temperature $T_m$.  One way to realize this is via e.g., a
two-component crystal, where high $T_m$ crystallization is controlled
by a strongly interacting atom and vacancy/interstitial core energy
$E_c$ by a weakly interacting smaller atom. One can view such system
in a limit of a positionally well-ordered single-component crystal,
determining $T_m$, acting as a periodic potential for a second
weakly-interacting atom, whose vacancy/interstitial core energies
$E_c \ll T_m$. This may be realizable in current highly tunable
ultracold-atom experiments.

\noindent{\em Reentrant superfluid:}
In fact, one can consider an extreme $T_m\rightarrow\infty$ limit with
a single atomic component in a commensurate periodic potential, that
is routinely realized in ultra-cold atoms via an optical potential of
interfering laser beams. Because of the imposed periodic potential
atoms are by construction in a crystal state, with the atomic
structure function exhibiting Bragg peaks at all temperatures. With
melting obstruction out of the way, the $T_{c1} \ll T_{c2}$
requirement may be satisfied by working at low, commensurate filling
fraction and short-range interactions, natural for neutral ultracold
atoms.

\begin{figure}[tbp]
  \includegraphics[height=2in,width=3in]{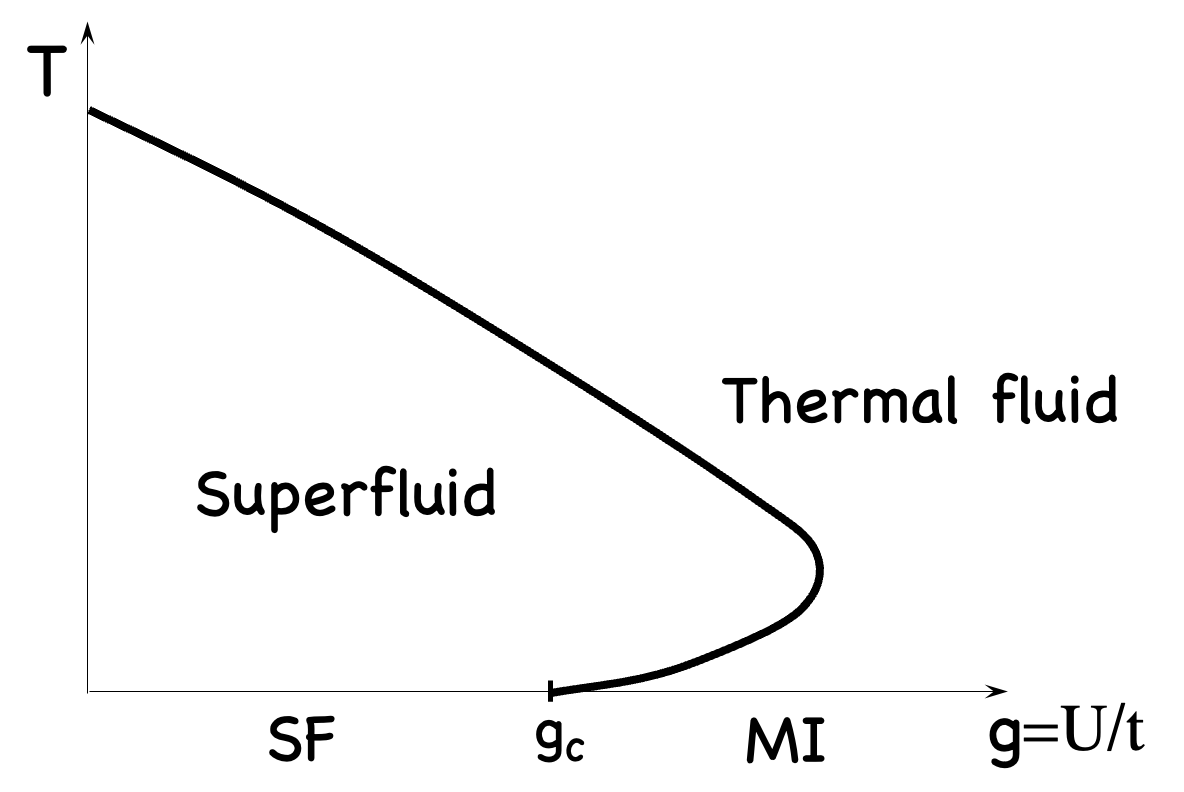}
  \caption{A schematic putative finite temperature MI-SF phase diagram showing a
    reentrant superfluid phase above a zero-temperature MI state.}
\label{figRentranceSF}
\end{figure}

However, one may then rightly object that the remaining problem has
nothing to do with supersolidity, since the translational symmetry is
{\em explicitly} broken by the imposed optical periodic
potential. Indeed this is correct, and the problem in this limit
instead reduces to a putative thermally-induced {\em reentrant
  superfluidity} of interacting lattice bosons, as I discuss below
and illustrate in Fig. \ref{figRentranceSF}.

At zero temperature and commensurate filling, such lattice bosons
exhibit an extensively studied Superfluid to Mott-insulator (SF-MI)
quantum phase transition\cite{Doniach81,FisherMISF89} as a function of
the ratio $U/t\equiv g$ of the onsite interaction strength $U$ to the
hopping amplitude $t$, realizable in ultracold
atoms\cite{Jaksch98,Greiner02}.  Focussing on the Mott-insulator side
of the quantum SF-MI critical point, I consider a nonzero temperature
at which a nonzero fraction of bosons $n(T) = n_0 e^{-E_c/T}$ is
thermally activated above the Mott gap $E_c$, whose scale is set by
$U$. This gas of bosons at nonzero temperature can then be in a
superfluid or a thermal state, depending on the ratio of its density
$n(T)$ and the deBroglie condensation density $n_c(T)$, latter
controlled by the hopping amplitude, $t$.
Thus, in this context, the scenario discussed above is then simply a
prediction of a possible (depending on microscopic details)
temperature-driven {\em reentrant} Normal-to-SF-to-Normal phase
transition of finite-temperature bosons on a lattice, as illustrated
in Fig.\ref{figRentranceSF}.  To realize such a scenario I note that
on one hand, to be in the zero-temperature MI state the system must
have $t < U$, but on the other hand, to have reentrance,
$T_{c1}\sim U < T_{c2}\sim t$, i.e., must have $U < t$.  These
considerations give a crude estimate of the phase boundaries
\bea
e^{-E_c/T} &\sim&  T/t\;,\label{reentSFa}\\
e^{-1/\hat T} &\approx& c g\hat T\;,\label{reentSFb}
\eea
(akin to that in the continuum, summarized by Eq. \rf{crossEq}), where
$\hat T \equiv T/E_c$ is the temperature in units of MI gap $E_c$,
that I expect to be proportional to $U$, $E_c/t\approx c U/t = c g$,
with $c$ the proportionality constant. Thus from \rf{reentSFb} we see
that the existence of the superfluid reentrance requires
$ c < g^{-1} < 1$, and is then a matter of microscopic details that
determine constants of $O(1)$, inaccessible to above generic arguments
and qualitative estimates. With the above motivation, a demonstration
of the reentrant superfluidity requires a detailed quantitative
analysis of the nonzero-temperature phase boundary that must include
next- and next-to-nearest neighbors interactions and hopping, and
faithfully treated thermal fluctuations.  With all the tunability
afforded by AMO systems, I expect this reentrant-superfluid scenario
to be realizable in a degenerate atomic Bose gas in an optical
periodic potential.\cite{Greiner02}

\noindent{\em Summary:}
In this brief note, I presented generic arguments for a possible
realization of a thermally-driven reentrant supersolid and superfluid.
The key to the mechanism is a strong low-temperature dependence of the
density of bosonic excitations, activated over a gap of the insulating
state such as a non-supersolid crystal and a Mott insulator. I
demonstrated that although such putative reentrance is in principle
possible it relies on favorable microscopic details, that lie beyond
present analysis. A role of interactions, and quantum and thermal
fluctuations will quantitatively modify above crude estimates of the
phase boundary.  Thus, a detailed first-principles numerical analysis
for a specific system is necessary to assess the feasibility of the
thermal reentrance scenario outlined here.  Despite this uncertainty,
the possibility of realizing a temperature-driven {\em reentrant}
supersolidity and superfluidity is quite intriguing, and I hope will
stimulate numerical and experimental searches of this phenomenology.

\emph{Acknowledgments}.  I thank Moses Chan, Anton Kapustin, Jorge
Kurchan, Alan Dorsey, Liang Fu, Victor Gurarie, Boris Svistunov and
Nikolay Prokof\'ev for feedback and stimulating discussions.  I
acknowledge support by the Simons Investigator Award from the James
Simons Foundation. I thank The Kavli Institute for Theoretical Physics
for its hospitality during workshops while this manuscript was in
preparation, supported by the National Science Foundation under Grant
No. NSF PHY-1748958 and PHY-2309135.

\end{document}